\documentclass[reprint,superscriptaddress,amsmath,amssymb,aps,prl]{revtex4-1}
\usepackage{graphicx}% Include figure files
\usepackage{placeins}
\usepackage{bm}% bold math
\usepackage{xcolor}

\begin{document}

\title{Multistable circular currents of polariton condensates trapped in ring potentials}

\author{Franziska Barkhausen}
\affiliation{Department of Physics and Center for Optoelectronics and Photonics Paderborn (CeOPP), Universit\"{a}t Paderborn, Warburger Strasse 100, 33098 Paderborn, Germany}

\author{Stefan Schumacher}
\affiliation{Department of Physics and Center for Optoelectronics and Photonics Paderborn (CeOPP), Universit\"{a}t Paderborn, Warburger Strasse 100, 33098 Paderborn, Germany}
\affiliation{College of Optical Sciences, University of Arizona, Tucson, AZ 85721, USA}

\author{Xuekai Ma}
\email{xuekai.ma@gmail.com}
\affiliation{Department of Physics and Center for Optoelectronics and Photonics Paderborn (CeOPP), Universit\"{a}t Paderborn, Warburger Strasse 100, 33098 Paderborn, Germany}

%% To be edited by editor
% \dates{Compiled \today}

%\ociscodes{(140.3490) Lasers, distributed feedback; (060.2420) Fibers, polarization-maintaining;(060.3735) Fiber Bragg gratings.}

%% To be edited by editor
% \doi{\url{http://dx.doi.org/10.1364/XX.XX.XXXXXX}}

\begin{abstract}
We demonstrate the formation and trapping of different stationary solutions, oscillatory solutions, and rotating solutions of a polariton condensate in a planar semiconductor microcavity with a built-in ring-shaped potential well. Multistable ring shaped solutions are trapped in shallow potential wells. These solutions have the same ring shaped density distribution but different topological charges, corresponding to different orbital angular momentum (OAM) of the emitted light. For stronger confinement potentials, besides the fundamental modes, higher excited (dipole) modes can also be trapped. If two modes are excited simultaneously, their beating produces a complex oscillation and rotation dynamics. When the two modes have the same OAM, a double-ring solution forms for which the density oscillates between the inner and the outer ring. When the two modes have different OAM, a rotating solution with a crescent-shaped density and fractional OAM is created.
\end{abstract}

\maketitle

\textit{Introduction} -- Optical vortices have been widely studied in nonlinear optics~\cite{swartzlander1992optical,malomed2001discrete,kartashov2005stable,yakimenko2005stable,brasselet2009optical,izdebskaya2018stable}. The vorticity, also known as topological charge, stores quantized phase information and can be carried losslessly by the vortex (soliton) during propagation. An optical vortex has a ring-shaped intensity profile and commonly the diameter of the vortex increases with the topological charge. Higher-order vortices with larger topological charges often become unstable during propagation~\cite{mamaev1997decay} but can be stabilized by saturable nonlinearity~\cite{soto1991stability}, nonlocal nonlinearity~\cite{briedis2005ring}, optical lattices~\cite{sakaguchi2005higher}, or localized gain in a dissipative system~\cite{lobanov2011stable}. 

For possible applications in information storage and processing, optically-induced excitations in planar semiconductor microcavities, so-called exciton-polaritons, have recently attracted a lot of attention. An exciton-polariton is a hybrid light and matter quasiparticle that is composed of quantum well (QW) excitons and cavity photons. Due to their half-light half-matter nature, polaritons can be probed and excited optically, both resonantly and nonresonantly, and they interact with each other through Coulomb interaction. The resulting polariton-polariton interaction gives rise to strong nonlinearities and spontaneous macroscopic coherence, i.e., polariton condensation~\cite{deng2002condensation,kasprzak2006bose}. The finite lifetime of polaritons renders the system inherently dissipative and the condensates emit coherent light by recombination of excitations. The (repulsive) nonlinearity of polaritons leads to manifestation of various nonlinear phenomena, such as optical bistability~\cite{baas2004optical,bajoni2008optical}, solitons~\cite{egorov2009bright,sich2012observation,ma2017creation}, and vortices~\cite{lagoudakis2008quantized,sanvitto2010persistent,gao2018controlled,ma2019realization}.

\begin{figure} [!b]
\includegraphics[width=1\columnwidth]{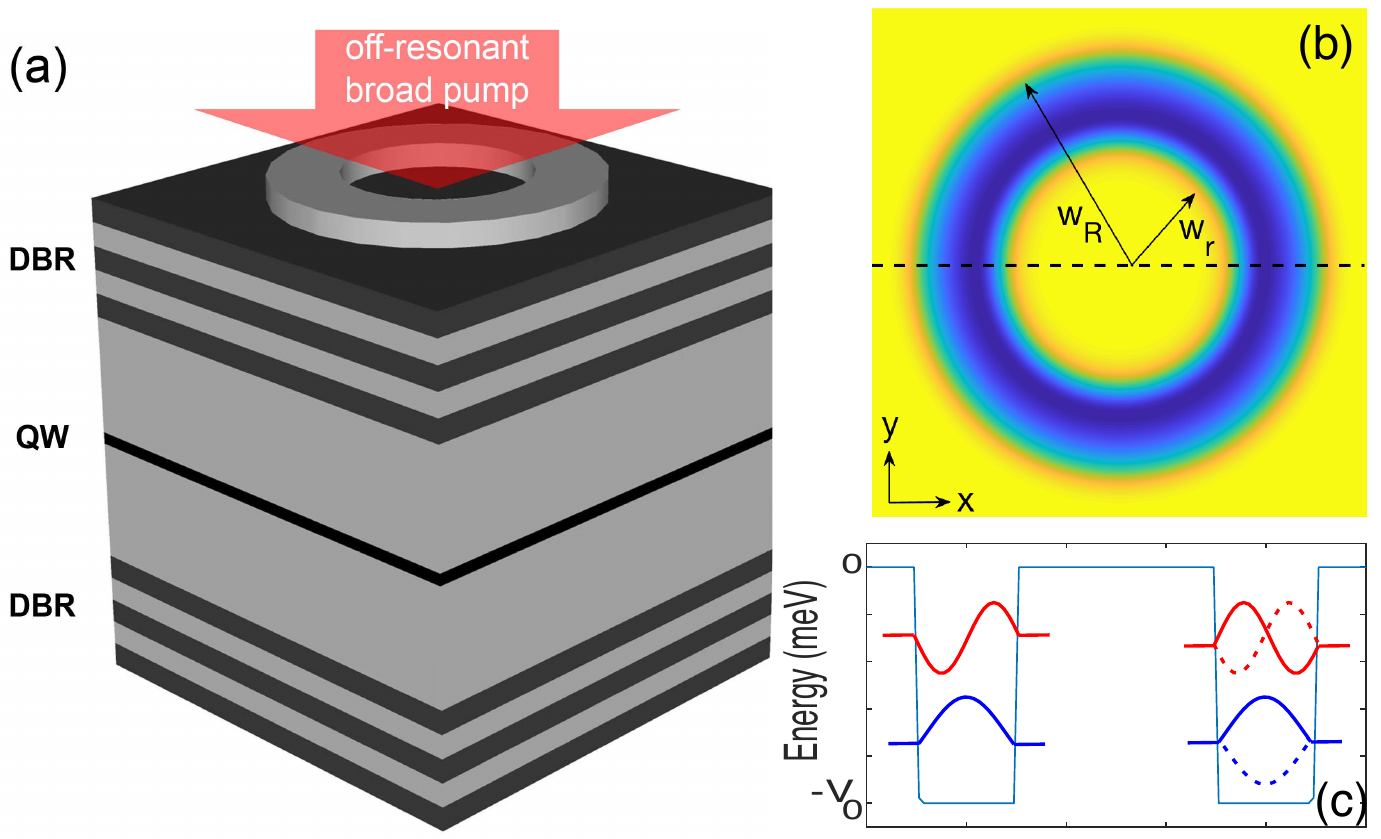}
\caption{(a) Sketch of a planar semiconductor microcavity containing a fabricated ring-shaped well potential. A broad non-resonant optical beam is applied for excitation. (b) In-plane landscape of the external potential. The width of the potential well is represented by the difference of the two radii, i.e. $w_R$-$w_r$. (c) 1D sketch of the potential along the dashed line in (b). Sketch of fundamental and dipole condensate modes trapped inside the wells. Dashed lines indicate the corresponding anti-symmetric states.}\label{sketch}
\end{figure}

Under nonresonant excitation, an optical beam induces the excitation reservoir that  provided the gain of the polariton condensate. In this case, vortices formed are balanced not only by the dispersion and nonlinearity (as in conservative systems), but also by the loss and the gain, such that they possess both OAM and a radial momentum. The excitation reservoir also acts as an effective external potential and strongly influences the distribution and dynamics of condensates. A ring-shaped optical beam can create a 2D potential trap, which balances the outgoing propagation of the condensate, supporting stable ring shaped (bright) vortices in a system with repulsive nonlinearity. Simultaneously, the feedback of the condensation process reshapes the distribution of the reservoir, enabling vortex multistability~\cite{ma2018vortex}.

In this paper, we consider a ring-shaped external well potential in a semiconductor microcavity as sketched in Fig.~\ref{sketch}(a). The potential well can trap the condensate inside, suppressing the outgoing propagation. In this case, multistable vortices can form that have (almost) the same ring shaped density profile but different OAM and energies. In recent works~\cite{dominici2018ultrafast,kartashov2019rotating} it was reported that two coherent optical beams with different OAM or frequencies can excite rotating condensates. Here we demonstrate that for nonresonant excitation, by tuning the depth and the width of the ring potential higher modes with two density rings inside the potential well can also be trapped. If two of these modes with different OAM contribute to the condensate simultaneously, spontaneously rotating solutions are obtained in the nonresonant excitation scenario where neither OAM nor frequency are externally imprinted. If the two modes have the same OAM, oscillating ring solutions are predicted. 

\textit{Model} -- The dynamics of a polariton condensate in a planar QW semiconductor microcavity in the vicinity of the polariton ground state can be described by a driven-dissipative Gross-Pitaevskii (GP) equation, coupled to an incoherent equation for the density of the excitation reservoir~\cite{wouters2007excitations}:
%\begin{widetext}
\begin{equation}\label{e1}
\begin{aligned}
i\hbar\frac{\partial\Psi(\mathbf{r},t)}{\partial t}&=\left[-\frac{\hbar^2}{2m_{\text{eff}}}\nabla_\bot^2-i\hbar\frac{\gamma_\text{c}}{2}+g_\text{c}|\Psi(\mathbf{r},t)|^2 \right.\\
&+\left.\left(g_\text{r}+i\hbar\frac{R}{2}\right)n(\mathbf{r},t)+V(\mathbf{r},t)\right]\Psi(\mathbf{r},t)
\end{aligned}
\end{equation}
\begin{equation}\label{e2}
\frac{\partial n(\mathbf{r},t)}{\partial t}=\left[-\gamma_r-R|\Psi(\mathbf{r},t)|^2\right]n(\mathbf{r},t)+P(\mathbf{r},t)\,.
\end{equation}
%\end{widetext}
Here $\Psi(\mathbf{r},t)$ is the coherent condensate field and $n(\mathbf{r},t)$ is the density of the reservoir. The effective mass of polaritons around the bottom of the lower-polariton branch is $m_{\text{eff}}{=}10^{-4}m_{\text{e}}$ ($m_{\text{e}}$ is the free electron mass). The loss rate of the condensate and the reservoir are represented by $\gamma_\text{c}{=}0.08~\mathrm{ps}^{-1}$ and $\gamma_\text{r}{=}1.5 \gamma_c$, respectively. $R{=}0.01~\mathrm{ps}^{-1}~\mu\mathrm{m}^2$ is the polariton condensation rate. The nonlinear coefficient $g_\text{c}{=}3\times 10^{-3}~\mathrm{meV}~\mu\mathrm{m}^2$ represents the strength of the polariton-polariton interaction and the strength of the polariton-reservoir interaction is given by $g_\text{r}{=}2 g_\text{c}$. $V(\mathbf{r},t)$ is the external potential. $P(\mathbf{r},t)$ is the nonresonant pump with frequency far above the exciton resonance. In this work, a Gaussian optical beam is used for excitation with spot diameter of $\sim$100 $\mu$m and intensity of $\sim$1.2$P_\text{th}$, where $P_\text{th}$ is the condensation threshold. The coupled equations, Eqs. (\ref{e1}) and (\ref{e2}), are explicitly solved using a 4th order Runge-Kutta method on a finite real-space grid with discretization of the dispersion with zero boundary conditions.

To create a trapped vortex, a ring-shaped external potential well is included in the system. Such kind of potential can be fabricated in planar semiconductor microcavities by different techniques~\cite{balili2007bose,lai2007coherent,wertz2010spontaneous,kim2013exciton,winkler2016collective}. The sketch of the ring potential is shown in Fig.~\ref{sketch}. The spatial distribution of the potential is shown in Fig.~\ref{sketch}(b), where $w_R$ and $w_r$ represent the radii of the outer and inner boundaries of the potential, so that the potential width is given by $w_R-w_r$. In this work, we vary the width of the potential by changing $w_R$ for $w_r=5\,\mu\text{m}$ fixed. The potential depth is $V_0$.

Inside the potential well in Fig.~\ref{sketch}(c), different vortex states can be excited including the fundamental modes, which refers to vortices with only one density ring, and dipole modes with two density rings. Higher modes can also be trapped if the potential becomes deeper and broader~\cite{roumpos2010gain,dreismann2014coupled}. With sufficient separation ($2w_r$) of the condensate on opposite sides of the ring, interference is significantly reduced. This enables the formation of multiple stationary solutions where both the fundamental and the dipole mode have two possible realizations, the $0$-state and the $\pi$-state. Considering a 1D cut through the ring as illustrated in  Fig.~\ref{sketch}(c), the $0$-state of the condensate has the same phase on both sides as illustrated by the solid blue (fundamental mode) and solid red (dipole mode) lines in Fig. \ref{sketch}(c). For the $\pi$-state the condensate wavefunction in the centrosymmetric potential well is anti-symmetric [solid lines in the left well and dashed lines in the right well in Fig.~\ref{sketch}(c)].

\begin{figure} %[htbp]
\includegraphics[width=1\columnwidth]{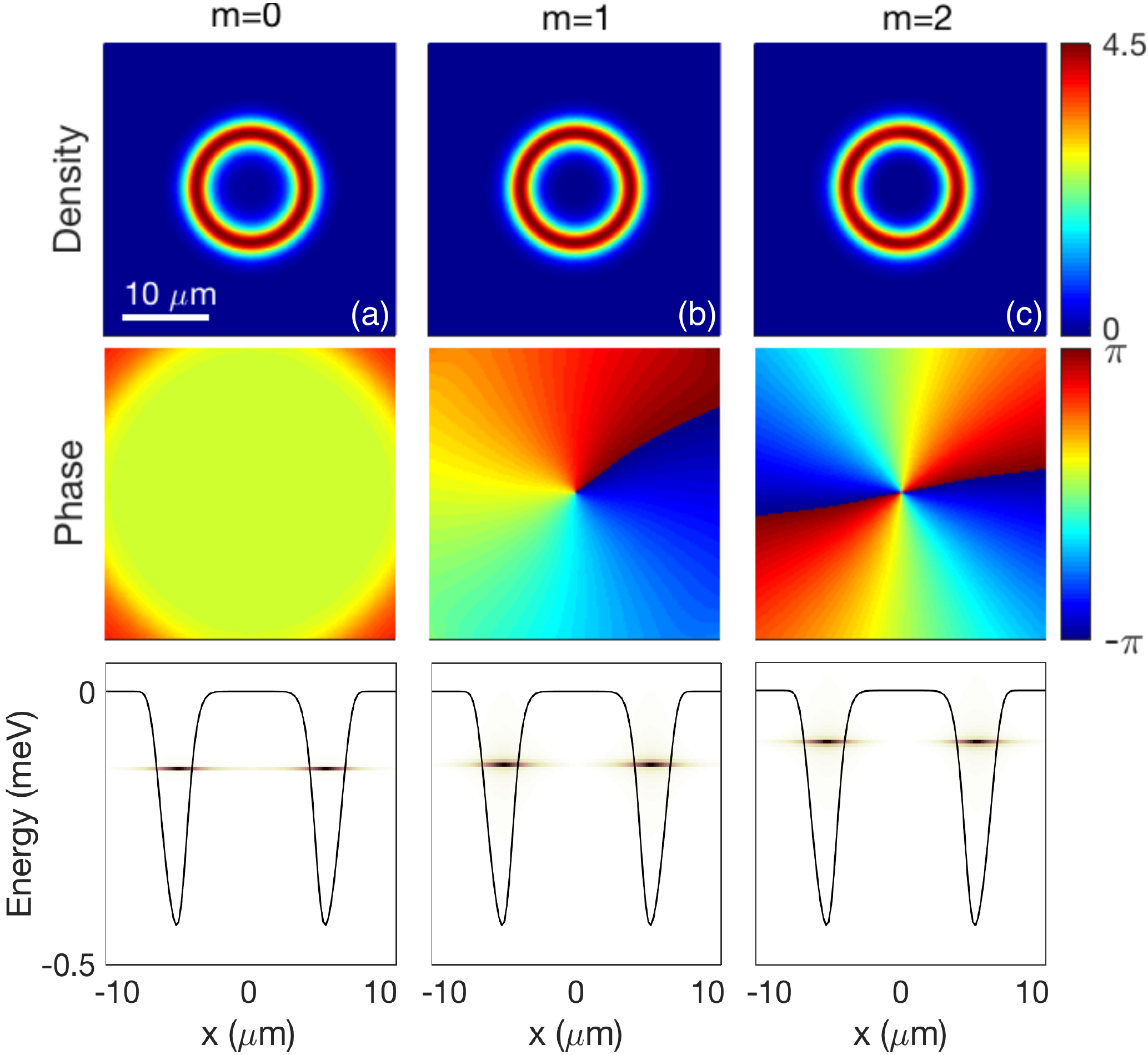}
\caption{{\bf Multistability of vortex states.} Distributions of the density (top row), the phase (middle row), and the spectra (bottom row) of multistable trapped vortices with topological charges (a) $m=0$, (b) $m=1$, and (c) $m=2$. The profile of the potential well with $w_R=7$ $\mu$m and $V_0=0.5$ meV is plotted on top of the spectra.}\label{multistability}
\end{figure}

\begin{figure} [h]
\includegraphics[width=1\columnwidth]{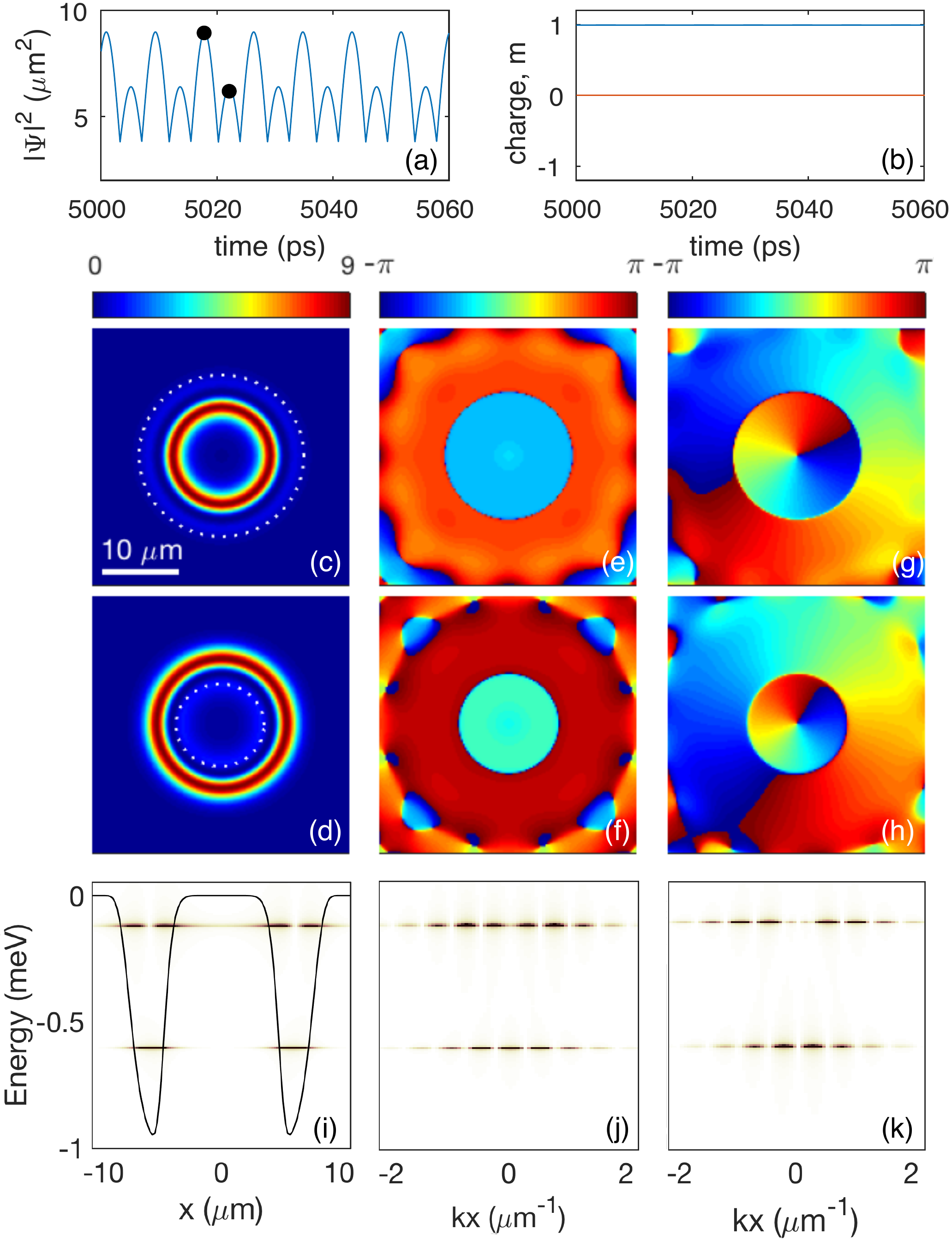}
\caption{{\bf Oscillating vortex states.} Time evolution of (a) the peak density and (b) the OAM of condensates in a broader ($w_R=8$ $\mu$m) and deeper ($V_0=1$ meV) potential well. The red line indicates the OAM of the states in (e,f), while the blue line indicates the OAM of the states in (g,h). Distributions of (c,d) density and (e,f) phase of the solutions at different times, corresponding to the black points in (a) with a time difference of $\Delta t=$ 4.3 ps. (g,h) Phase distributions of oscillating vortices with $m=1$. Their density distributions are the same with those in (c,d). (i) Real-space spectrum for the solution in (e,f). (j) $k-$space spectrum of the solution in (e,f). (k) $k-$space spectrum of the solution in (g,h). The well radii are marked by the dashed circles in (c,d). The potential well is included in (i).}\label{oscillation}
\end{figure}

First, we consider the $0-$ and $\pi-$states of the fundamental mode. For the $0-$state, if traveling to the opposite side of the ring, the phase difference must be $2\pi\cdot n$, with integer $n=0,\ 1,\ 2,\ \cdots$. Therefore the phase difference, $\Theta$, of a vortex solution by travelling the whole ring satisfies
\begin{equation}\label{TopoChar}
\frac{\Theta}{2}=2\pi\cdot n, \ \ \ n=0,\ 1,\ 2,\ \cdots
\end{equation}
In this case, a trapped ring solution can carry, in principle, the topological charge $m=\frac{\Theta}{2\pi}=0,\ 2,\ 4,\ \cdots$. Similarly, it can be calculated that for the $\pi-$state solutions the topological charges can be $m=1,\ 3,\ 5,\ \cdots$. The non-topological ($m=0$) ring solution and the $m=1$ and $m=2$ charged vortex solutions are presented in Fig. \ref{multistability}. All these solutions are stable after a sufficient long time evolution. To target the solutions individually, a seed with the respective OAM is used in the numerical solution of Eqs.~(1) and (2). One can see from the condensate spectra (the bottom row in Fig. \ref{multistability}) that the solutions are completely trapped inside the potential well, so that their density distributions (the top row in Fig. \ref{multistability}) are almost the same with perfectly localized rings. This property is significantly different from the multistable vortices formed for ring shaped pump profiles without external potential~\cite{ma2018vortex}, where the effective diameter of a vortex ring increases with the topological charge. Another difference is the phase distribution. In previous work the phases of polariton vortices are spiraling due to the outgoing propagation of the condensate~\cite{ma2016incoherent,ma2018vortex}, while in this work the strong and narrow potential well leads to propagation of the condensate only inside the well. As a result, the phases are almost radius independent (middle row in Fig. \ref{multistability}). This enables the stabilization of higher order vortices that are topologically unstable in optical potentials, with the density minimum undergoing dynamical splitting and recombination~\cite{dall2014creation}, in polariton condensates. The vortex with $m=\pm3$ ($\pi-$state) is unstable for the same parameters. The higher the topogical charge, the higher the vortex energy, such that the $m=\pm3$ vortex is nearly above the potential well. We note that after sufficiently long time evolution the $m=\pm3$ vortex transitions into the fundamental mode with $m=\pm1$, which is the $\pi-$state, instead of the fundamental solution with $m=0$ (0-state). Higher topological charges can be trapped and stabilized if the potential depth or the width is increased. 

Figure~\ref{oscillation} shows the simultaneous appearance of the fundamental mode and the dipole mode when the potential becomes deeper and broader. The beating between the two trapped modes creates an oscillating solution with two different peak densities as shown in Fig.~\ref{oscillation}(a). The period is around $8.6\,\mathrm{ps}$, which agrees with the energy difference of the two states in Fig.~\ref{oscillation}(i). The dipole mode is a higher mode in radius direction, with two density rings, one close to the inner boundary of the potential well and the other one close to the outer boundary. Superimposing this higher mode with the fundamental mode results in a density that is persistently oscillating between the two rings as shown in Figs. \ref{oscillation}(c) and \ref{oscillation}(d). The topological charge of such an oscillation solution is time independent with $m=0$. This solution is generated for a homogeneous initial condition with a flat phase profile.  

\begin{figure} [!b]
\includegraphics[width=1\columnwidth]{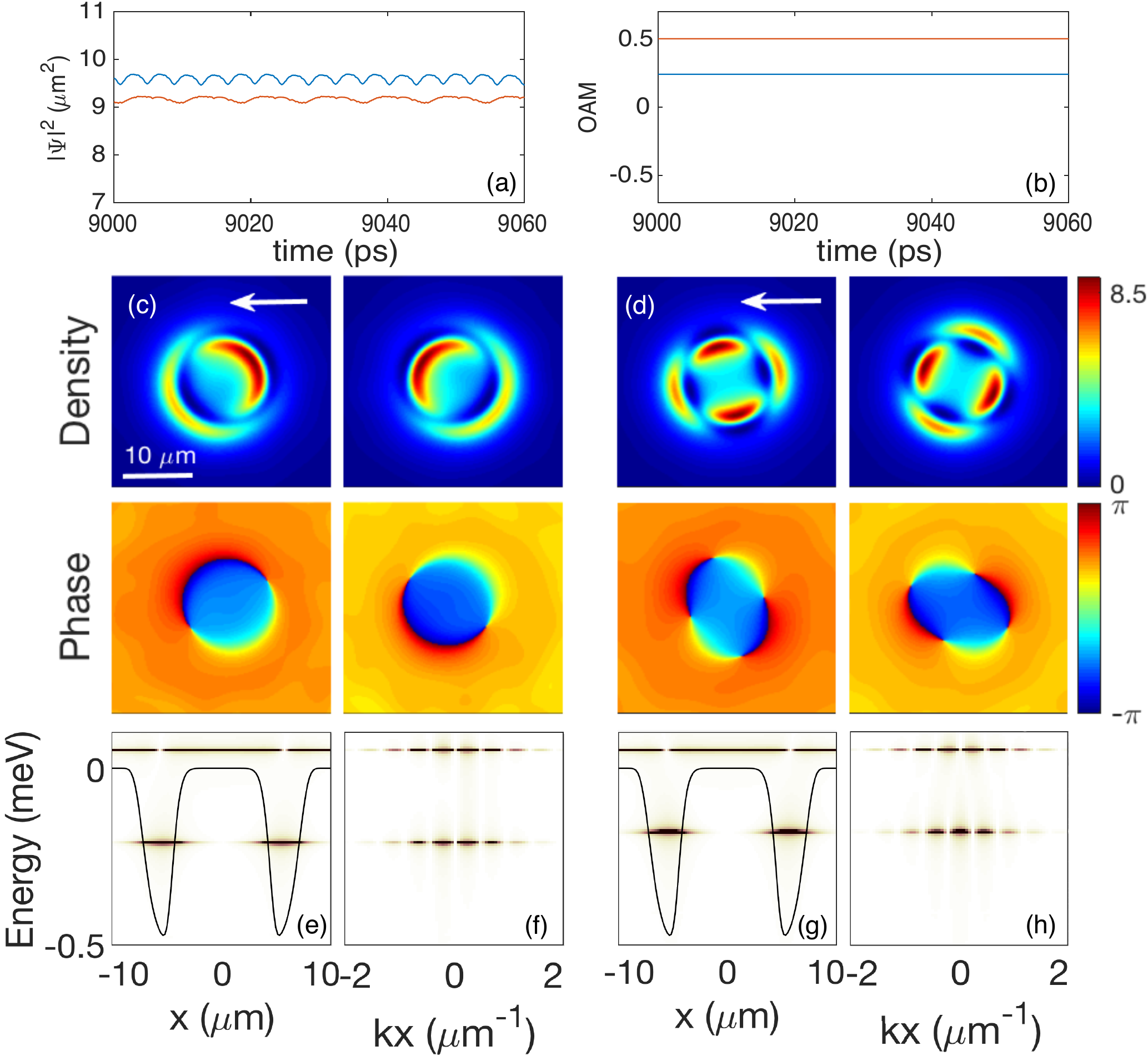}
\caption{{\bf Rotating vortex states.} Time evolution of (a) the peak density and (b) the OAM of condensates in the  potential well with $V_0=0.5$ meV and $w_R=8$ $\mu$m. The blue lines correspond to the solution in (c) and the red lines correspond to the solution in (d). (c,d) Distributions of density (upper row) and phase (lower row) of condensates at different time scales: $\Delta t=$ 4.5 ps for (c) and $\Delta t=$ 4.5 ps for (d). The arrows represent the rotation directions, which are decided by the initial conditions. Spectra in (e) real space and (f) $k-$space of the solution in (c). Spectra in (g) real space and (h) $k-$space of the solution in (d). The potential well is included in (e,g).}\label{rotation}
\end{figure}

If the initial condition carries an OAM close to $m=1$, the same peak density oscillation with that in Fig. \ref{oscillation}(a) is obtained due to the simultaneous contribution of the two modes. But, in this case the topological charge of the condensate converges to $m=1$ [blue line in Fig.~\ref{oscillation}(b)]. The $2\pi$ phase jump of the ring solution is clearly seen in Figs.~\ref{oscillation}(g) and \ref{oscillation}(h), where a $\pi$ phase jump between the outer ring and the inner ring is seen. We note that with the similar oscillation dynamics in real space there is no significant difference between the momentum-resolved spectra of the two oscillations as shown in Fig.~\ref{oscillation}(i). If comparing their spectra in $k$ space, one can see that in Fig.~\ref{oscillation}(j) both fundamental mode and higher mode are excited to the $0$-state without OAM with a contribution to the spectrum at $k_x=0$. In contrast in Fig.~\ref{oscillation}(k) two $\pi$-state solutions, i.e. fundamental vortices with $m=1$, are excited.

%\begin{figure} %[htbp]
%\includegraphics[width=1\columnwidth]{oscillation.pdf}
%\caption{Time evolutions of (a) the peak density and (e) the phase ($m=1$) of condensates in a broader ($w_R=8$ $\mu$m) and deeper ($V_0=-0.1$ meV) potential. Distributions of (b,f) density and (c,g) phase of the solutions at different time scales, corresponding to the black points in (a), respectively. Spectra in (d) real space and (h) $k$ space of the solution in (a). The spatial distribution of the potential is marked by the dashed circles in (b,f). The 1D potential well is placed in (d).}\label{oscillation}
%\end{figure}

In the results shown above, the strongly trapped fundamental and dipole modes always inherit the phase from the initial conditions. Therefore, it is worth considering a scenario where these two modes are in opposite states. For this purpose, we use noisy initial conditions and slightly decrease the depth of the ring potential to $V_0=0.5$ meV, with the same width $w_R=8$ $\mu$m as in Fig.~\ref{oscillation}. The shallower potential squeezes the dipole mode towards potential edge, forming a free (non-trapped) mode when it is out of the potential well as shown in Fig.~\ref{rotation}(e). The k-resolved spectra in Figs.~\ref{rotation}(e) and (f) show that the higher dipole mode is the $0$-phase state (with $m=0$), while the fundamental mode is still the $\pi$-state (with $m=1$). Since their is only one mode that is topologically charged, the simultaneous excitation of the two modes shows a crescent shaped density profile in both the inner ring and the outer ring in Fig.~\ref{rotation}(c). This density distribution preserves its shape over time and is rotating about the center of the ring. Such dynamics lead to a total OAM that takes a fractional value as shown Fig.~\ref{rotation}(b)]. Interestingly, the fundamental mode can also be excited to the $m=2$ vortex (0-state) as shown in Figs.~\ref{rotation}(g) and (h) due to the multistability already shown in Fig.~\ref{multistability}. In this case, coupling with the $m=0$ dipole mode results in a rotating wheel pattern, carrying also a fractional OAM as shown in Fig.~\ref{rotation}(b).

To conclude, we have demonstrated that polariton condensates trapped in an external ring-shaped well potential show a multistability of fundamental states with or without OAM. When the potential is sufficiently deep and broad both fundamental and dipole modes can be excited. Then, depending on the OAM of the dominant modes, a condensate state is generated that is oscillating between an inner and an outer ring or forming a spatially rotating state with crescent shape.

This work was supported by the Deutsche Forschungsgemeinschaft (DFG) through the collaborative research center TRR142 (grant No. 231447078, project A04) and Heisenberg program (grant No. 270619725) and by the Paderborn Center for Parallel Computing, PC$^2$. X.M. further acknowledges support from the National Natural Science Foundation of China (Grant No. 11804064).

%Bibliography
%\bibliography{sample}

% Full bibliography added automatically for Optics Letters submissions; the following line will simply be ignored if submitting to other journals.
% Note that this extra page will not count against page length

%\bibliographystyle{unsrt}
%\bibliography{sample}
%Manual citation list
%\begin{thebibliography}{1}
%\bibitem{Zhang:14}
%Y.~Zhang, S.~Qiao, L.~Sun, Q.~W. Shi, W.~Huang, %L.~Li, and Z.~Yang,
 % \enquote{Photoinduced active terahertz metamaterials with nanostructured
  %vanadium dioxide film deposited by sol-gel method,} Opt. Express \textbf{22},
  %11070--11078 (2014).
%\end{thebibliography}

\end{document}